\documentstyle[11pt]{article}

\newcommand{\beq}{\begin{equation}}
\newcommand{\eeq}{\end{equation}}
\newcommand{\beqa}{\begin{eqnarray}}
\newcommand{\eeqa}{\end{eqnarray}}

\newcommand{\NPB}[1]{{\it Nucl. Phys.}\ {\bf B{#1}}}
\newcommand{\PLB}[1]{{\it Phys. Lett.}\ {\bf B{#1}}}
\newcommand{\PRD}[1]{{\it Phys. Rev.}\ {\bf D{#1}}}
\newcommand{\PRL}[1]{{\it Phys. Rev. Lett.}\ {\bf #1}}

\def\st{\sin\theta}
\def\ct{\cos\theta}
\def\sp{\sin\phi}
\def\cp{\cos\phi}

\def\gp{g^\prime}
\def\gq{g_l}
\def\gl{g_h}

\begin{document}
\begin{titlepage}
\def\thepage {}        

\title{Limits on Non-Commuting Extended Technicolor}

\author{
R.S. Chivukula\thanks{e-mail addresses: sekhar@bu.edu,
simmons@bu.edu, terning@calvin.bu.edu},
E.H. Simmons, and J. Terning\\
Department of Physics, Boston University, \\
590 Commonwealth Ave., Boston MA  02215}

\date{June 26, 1995}

\maketitle

\bigskip
\begin{picture}(0,0)(0,0)
\put(295,250){BUHEP-95-19}
\put(295,235){hep-ph/9506427}
\end{picture}
\vspace{24pt}

\begin{abstract}
Using precision electroweak data, we put limits on
non-commuting extended technicolor models.  We conclude that these
models are viable only if the ETC-interactions are strong.
Interestingly, these models predict a pattern of deviations from the
standard model which can fit the data significantly better than the
standard model does, even after taking into account the extra
parameters involved.
\pagestyle{empty}
\end{abstract}
\end{titlepage}

\section{Introduction}
\label{sec:intro}

There continue to be several small discrepancies between precision
electroweak measurements and the predictions of the standard model
\cite{langacker,blondel,langackerbb}.  The most interesting is
associated with the ratio of the $Z$ decay widths to $b
\overline{b}$ and to all hadrons ($R_b$).  In addition if
$\alpha_s(M_Z) \approx 0.115$, as suggested by recent lattice results
\cite{lattice} and deep-inelastic scattering \cite{langacker,deep},
there are also potentially significant deviations in the ratios of the
hadronic to leptonic widths.  If new physics which increased $R_b$
were present, a better fit to precision electroweak data would be
obtained with a value of $\alpha_s(M_Z)$ comparable to the value
quoted above \cite{alphaZbb}.

Recently it was shown that in non-commuting extended technicolor (ETC)
models, in which in which the ETC interactions \cite{ETC} do not
commute with the $SU(2)_L$ interactions of the standard model (i.e.,
in which $SU(2)_L$ is partially embedded in the ETC gauge group),
$R_b$ could exceed the standard model value \cite{NCETC}.  In that
discussion, we assumed that only the $(t, b)$
doublet had non-commuting ETC couplings suppressed by a scale low
enough to provide observable consequences at present machines.  {}From
the point of view of anomaly cancellation, it is much more economical
to assume that the entire third generation has the same non-commuting
ETC interactions. Such a scheme implies that the electroweak
interactions of the $\tau$ and $\nu_\tau$ will also exhibit
interesting deviations.

In this paper we will use the wealth of precision electroweak data to
place constraints on such non-commuting ETC family models. After
reviewing non-commuting ETC models, we discuss the deviations from the
standard model that this new physics would produce.
We discuss the amount of fine-tuning these models require to accommodate the
top quark mass and to agree with the precision electroweak data.
We conclude that these models predict a pattern of deviations from the
standard model which can fit the data significantly better than the
standard model does, even after taking into account the extra
parameters involved.

\section{Non-Commuting Extended Technicolor}
\label{sec:topcolor}
\setcounter{equation}{0}

\subsection{Gauge Symmetry-Breaking Pattern}

The pattern of gauge symmetry breaking that is required in
non-commuting ETC models is more complicated than that
in  commuting ETC models; it generally involves
three scales (rather than just two) to provide masses for one family
of ordinary fermions.  The required pattern of breaking is as
follows:
\begin{center}
$G_{ETC}  \otimes SU(2)_{light} \otimes U(1)' $
\end{center}
\vspace{-20pt}
\begin{center}
$\downarrow \ \ \ \ \ f $
\end{center}
\vspace{-20pt}
\begin{center}
$G_{TC} \otimes SU(2)_{heavy}  \otimes SU(2)_{light} \otimes U(1)_Y $
\end{center}
\vspace{-20pt}
\begin{center}
$\downarrow\ \ \ \ \ u $
\end{center}
\vspace{-20pt}
\begin{center}
$G_{TC}  \otimes SU(2)_{L} \otimes U(1)_Y$
\end{center}
\vspace{-20pt}
\begin{center}
$\downarrow\ \ \ \ \ v $
\end{center}
\vspace{-20pt}
\begin{center}
$G_{TC}  \otimes U(1)_{em}$,
\end{center}
The ETC gauge group is broken to technicolor and an
$SU(2)_{heavy}$ subgroup at the scale $f$.  The $SU(2)_{heavy}$ gauge
group is effectively the weak gauge group for the third generation in
these non-commuting ETC models, while the $SU(2)_{light}$ is the weak
gauge group for the two light generations.  The two $SU(2)$'s are
mixed (i.e. they break down to a diagonal $SU(2)_L$ subgroup) at the
scale $u$.  Finally the electroweak gauge symmetry breaking
is accomplished at the scale $v$, as is standard in technicolor
theories.

The two simplest possibilities for the $SU(2)_{heavy} \times
SU(2)_{light}$ transformation properties of the order
parameters that produce the correct combination of mixing and
breaking of these gauge groups are:
\beq
\langle \varphi \rangle \sim (2,1)_{1/2},\ \ \ \ \langle
\sigma\rangle \sim (2,2)_0 ~,\ \ \ \ \ \ \ ``{\rm heavy\ case}"~,
\eeq
and
\beq
\langle \varphi \rangle \sim (1,2)_{1/2},\ \ \ \ \langle
\sigma\rangle \sim (2,2)_0 ~,\ \ \ \ \ \ \ ``{\rm light\ case}"~.
\eeq
Here the order parameter $\langle\varphi\rangle$ is responsible for
breaking $SU(2)_L$ while $\langle\sigma\rangle$ mixes
$SU(2)_{heavy}$ with $SU(2)_{light}$.  We refer to these two
possibilities as ``heavy'' and ``light'' according to whether
$\langle\varphi\rangle$
transforms non-trivially under $SU(2)_{heavy}$ or $SU(2)_{light}$.

The heavy case, in which $\langle\varphi\rangle$ couples to the heavy
group, is the choice made in \cite{NCETC}, and corresponds to the case
in which the technifermion condensation responsible for providing mass
for the third generation of quarks and leptons is also responsible for
the bulk of electroweak symmetry breaking (as measured by the
contribution made to the $W$ and $Z$ masses).  The light case, in which
$\langle\varphi\rangle$ couples to the light group, corresponds to the
opposite scenario: here the physics responsible for providing mass for the
third generation {\it does not} provide the bulk of electroweak
symmetry breaking. While this light case is counter-intuitive (after
all, the third generation is the heaviest!), it may in fact provide a
resolution to the issue of how large isospin breaking can exist in the
fermion mass spectrum (and, hence, the technifermion spectrum)
without leaking into the $W$ and $Z$ masses.
This is essentially what happens in multiscale models
\cite{multi,chiraltc} and in top-color assisted technicolor
\cite{tcii}.  Such hierarchies of technifermion masses are also useful
for reducing the predicted value of $S$ in technicolor
models\footnote{Recently the experimental upper bound on $S$ has been
relaxed, so that substantially positive values of $S$ are allowed.
Ref. \cite{tc2isospin} gives $S < 0.4$ at the 95\% confidence level.}
\cite{revenge}.

\subsection{Top Quark Mass Generation}

In ETC models the top quark mass \cite{ETC} is generated by
four-fermion operators induced by ETC gauge boson exchange.  In
non-commuting ETC models, the left-handed third generation quarks and
right-handed technifermions, $\psi_L = (t,b)_L$ and $T_R = (U,D)_R$,
are doublets under $SU(2)_{heavy}$ while the left-handed
technifermions are $SU(2)_{heavy}$ singlets, and these operators may
be written as
\beq
{\cal L}_{4f} = - {2\over f^2}\left(\xi\bar\psi_L \gamma^\mu U_L +
{1\over \xi}\bar t_R \gamma^\mu T_R\right) \left(\xi \bar U_L \gamma_\mu
\psi_L + {1\over \xi}\bar T_R \gamma_\mu t_R \right) ~,
\label{4f}
\eeq
where $\xi$ is a model-dependent Clebsch.  When the technifermions
condense the $LR$ cross-terms in the operator (\ref{4f}) produce a
top quark mass.  In strong-ETC models, i.e.  models in which the ETC
coupling is fine-tuned to be close to the critical value necessary for
the ETC interactions to produce chiral symmetry breaking, the
$LR$-interactions become enhanced.  Physically, this is due to the
presence of a composite scalar \cite{Comp} which is light compared to
$f$ and communicates electroweak symmetry breaking to the top quark.

Whether or not the ETC interactions are strong, we can write the top
quark mass as \cite{ETC}
\beq
m_t\approx {g^2 4\pi f_Q^3 \over M^2} ~,
\label{mt}
\eeq
where the numerator contains an estimate of the technifermion
condensate (using dimensional analysis \cite{dimanal}), and $f_Q$ is
the Goldstone boson decay constant associated with the technifermions
which feed down a mass to the top quark.  In the heavy case the
technifermions responsible for giving rise to the third-generation
masses also provide the bulk of the $W$ and $Z$ masses, and we expect
$f_Q \approx 125$ GeV (which, for $m_t \approx 175$ GeV, implies $M/g
\approx 375$ GeV).  Even in the light case there must be some
$SU(2)_{heavy}$ breaking vev, that is $f_Q \neq 0$, in order to give
the top-quark a mass.  However, it is unreasonable to expect that
$M/g$ can be tuned to be smaller than $f_Q$ and equation (\ref{mt})
implies the lower bound $f_Q > 14$ GeV.

In an ETC theory with no fine-tuning, $M$ represents the
mass of the ETC gauge boson, which
is related to $f$ by
\beq
M  = {g f \over 2}~,
\eeq
where $g$ is the ETC gauge coupling.  In strong-ETC theories
\cite{strongETC}, $g^2$ is
the product of the scalar couplings to the top quark and the
technifermions and $M$ is the mass of the light composite
scalar \cite{Comp}.  In
the context of these theories, the ``accuracy'' with which the ETC
gauge coupling must be adjusted is approximately equal to the ratio of
$M^2/g^2$ to its naive value, $f^2/4$. That is, a rough measure of
fine-tuning required is
\beq
{4 M^2 \over g^2 f^2} \approx {8\pi f^3_Q \over m_t f^2}.
\label{fine}
\eeq

\subsection{Shifts in $Z$ Couplings}

As discussed in Ref. \cite{NCETC}, the $LL$ (and, in principle, $RR$)
terms in the operator (\ref{4f}) produce a shift in the $Z b
\overline{b}$ coupling:
\beq
\delta g^b_L = -{\xi^2 f_Q^2 \over { 2 f^2}} ~.
\label{dgLb}
\eeq
(For convenience, we have factored out $e/ \st \ct $ from all the $Z$
couplings.)
The difference between the scenario proposed in Ref. \cite{NCETC}
and that proposed here is that here the entire third family couples to the same
ETC gauge bosons, so there will also be corrections for the $\tau$ and
$\nu_\tau$ couplings:
\beq
\delta g^\tau_L = \delta g^{\nu_\tau}_L =
- {\xi^2 f_L^2 \over { 2 f^2}} ~,
\eeq
where we have allowed for the possibility that the technilepton
condensate is different from the techniquark condensate (i.e.
$f_Q \ne f_L$).

{}From the analysis given in Ref. \cite{NCETC}, we see that {\it in the
absence of fine-tuning} we expect the four-fermion operator (\ref{4f})
to induce corrections to the third-generation couplings of the order
of a few percent. Our fits to precision electroweak measurements will
allow us to put a lower bound on the size of $f$. We will translate
this, using equation (\ref{fine}) into an estimate of the amount of
fine-tuning required to produce a viable theory.

\section{Weak Boson Mixing: Heavy Case}
\label{sec:heavy}
\setcounter{equation}{0}

The remaining corrections come from weak gauge boson mixing\footnote{
The discussion of gauge boson mixing presented in this section and the
next follows the discussion of mixing in the un-unified
standard model given in Refs. \protect\cite{ununify,unun}.}. The $U(1)_{em}$
to which the electroweak group breaks is generated by
\beq
Q= T_{3l}+T_{3h}+Y~.
\label{q}
\eeq
The photon eigenstate can be written in terms of two weak mixing
angles,
\beq
A^\mu = \st \sp \,W_{3l}^\mu + \st \cp \,W_{3h}^\mu +\ct X^\mu~,
\label{pho}
\eeq
where $\theta$ is the weak angle and $\phi$ is an additional mixing angle.
Equations (\ref{q}) and (\ref{pho}) imply that the gauge couplings are
\beqa
\gq &=& {e\over s\st}~, \nonumber \\
\gl &=& {e\over c\st}~,\nonumber \\
\gp &=& {e\over \ct}~,
\eeqa
where $s\equiv\sp\,$ and $c\equiv\cp$.

Consider the low energy weak interactions in the four-fermion
approximation.  For the charged-currents, if we combine the
left-handed light and heavy fermion currents into the vector
$J^\dagger= (j_l, j_h)$, then the four-fermion interactions are
$2\,J^\dagger\,V^{-2}\,J$ where
\beq
V^{-2}={1\over u^2\,v^2}\,\pmatrix{
u^2+v^2&u^2\cr u^2&u^2}~.
\eeq
Retaining the standard relation between the $\mu$ decay rate and $G_F$
requires
\beq
\sqrt{2} G_F = {1\over v^2} + {1\over u^2}~.
\label{gfid}
\eeq
Hence the charged-current four-fermion weak interactions can be written
\beq
2\sqrt{2} G_F\left( j_l + j_h \right)^2 - {2 \over
u^2}(j_h^2 + 2 j_l j_h\,)~.
\label{wcc}
\eeq
The extra $j_l j_h$ term will affect the weak decays of
third-generation fermions.  For example, the effect on the tau decay
rate to muons is illustrated in Appendix A (equation (A.22)).  The
low-energy neutral-current interactions can be obtained similarly in
terms of $j_{em}$, the electromagnetic fermion current, and
$j_{3l,3h}$, the left-handed $T_3$ currents of light- and
heavy-charged fermions.  Applying (\ref{gfid}) gives
\beqa
 & & 2\sqrt{2} G_F \, (j_{3l} +
j_{3h} -\, j_{em}\sin^2\theta)^2  \\ \nonumber
 & - & {2\over \,u^2}\, \left[-j_{3h}^2 - 2 j_{3l} j_{3h} + \sin^2\theta\
j_{em}(2
j_{3h} + 2 j_{3l} c^2 + j_{em}\sin^2\theta (s^4 - 1))\right]~.
\label{ncplus}
\eeqa
Note that the extra neutral-current interactions involving third
generation fermions are suppressed only by $v^2/u^2$, whereas those
involving only charged first and second generation fermions are
additionally suppressed by mixing angles.

 It is convenient to discuss the mass eigenstates in the rotated basis
\beqa
W^{\pm}_1 &=& s\,W^{\pm}_l+c\,W^{\pm}_h\,, \\
W^{\pm}_2 &=& c\,W^{\pm}_l-s\,W^{\pm}_h\,, \\
Z_1 &=&\ct\,(s\,W_{3l}+c\,W_{3h})-\st\,X\,, \\
Z_2 &=& c\,W_{3l}-s\,W_{3h}\,,
\eeqa
in which the gauge covariant derivatives separate neatly into standard and
non-standard pieces
\beqa
D^\mu &=&\partial^\mu + ig\left( T_l^\pm + T_h^\pm \right) W^{\pm\,\mu}_1
+ig\left( {c \over s}T_l^\pm - {s \over c}T_h^\pm \right) W^{\pm\,\mu}_2
\nonumber \\
& &  +\  i{g \over {\ct}}\left( T_{3l} + T_{3h} -
\sin^2\theta \,Q \right) Z^\mu_1 +
ig\left( {c \over s}T_{3l} - {s \over c}T_{3h} \right)
Z^\mu_2.
\eeqa
where $g\equiv {e \over \st}$.
The breaking of $SU(2)_L$ results in mixing of  $Z_1$ and $Z_2$, as
well as a mixing of $W^{\pm}_1$ and $W^{\pm}_2$.
The mass-squared matrix for the $Z_1$ and $Z_2$
is:
\beq
M_Z^2=\left({e v \over {2 \st}} \right)^2\,
\pmatrix{{1\over \cos^2\theta}&
{- s\over c\cos\theta}\cr{- s\over c\cos\theta}&
{x\over s^2c^2}+{s^2\over c^2}\cr}\,,
\eeq
where $x= u^2 / v^2 $.  The mass-squared matrix for $W_1$ and $W_2$ is
obtained by setting $\ct = 1$ in the above matrix.

Diagonalizing the $W$ and $Z$ mass matrices in the limit of large $x$
and taking into account the value of $G_F$ given in equation (\ref{gfid}),
we find for the lightest eigenstates
\beq
M_W^2 \approx
\left({\pi \alpha_{em} \over{\sqrt{2} G_F \sin^2\theta}}\right)
\left(1 + {1\over x}(1 - s^4)\right),
\eeq
\beq
W^L\approx W_1+{c s^3 \over {x}}\,W_2
\eeq
and
\beq
M_Z^2 \approx
\left({\pi \alpha_{em} \over{\sqrt{2} G_F \sin^2\theta\cos^2\theta}}\right)
\left(1 + {1\over x}(1 - s^4)\right),
\label{dmzgf}
\eeq
\beq
Z^L\approx Z_1+{c s^3 \over {x\cos\theta}}\,Z_2~.
\label{dgz}
\eeq
Note that to this order in $1/x$, the custodial isospin relation for the $W$
and
$Z$ masses is preserved.

Thus, weak gauge boson mixing produces a shift from the standard model $Z$
couplings:
\beq
\delta g^f_L =  {{c s^3}\over x}
\left({c \over s}\, T_{3l} - {s \over c} \,T_{3h}\right)~.
\eeq
Furthermore, the relationship between $M_Z$, $G_F$, and $\sin^2\theta$
given in equation (\ref{dmzgf}) differs from that in the standard
model\footnote{ In the notation of Ref. \protect\cite{fit}, we can
account for this effect by the replacement $\Delta_e + \Delta_\mu =
(1-s^4)/x$.}.   Both effects must be taken into account in comparing
the predictions of this model with those of the standard model.

The result of all these corrections is that the predicted values of
many electroweak observables are altered from those given by the
standard model\footnote{We are using $\alpha_{em}(M_Z)$, $G_F$, and
$M_Z$ as the tree-level inputs.} \cite{unun,fit}.  For example, we
find that the $W$ mass is changed as follows:
\begin{equation}
 M_W = \left( M_W \right)_{SM} \left(1
-0.213 \left(1 - s^4\right)  {1 \over x} \right) ~.
\end{equation}
Likewise, one finds that the total width of the $Z$ becomes:
\beqa
 \Gamma_Z &=& \left( \Gamma_Z \right)_{SM} \left(1 -0.707 \delta g_L^b
-0.144 \delta g_L^\tau + 0.268 \delta g_L^{\nu_\tau}\right.\nonumber\\
 & &\left.\,\,\,\,\,\,\,\,\,\,\,\,\,\,\,\,\,\,\,\,\,\,\,\,\,\,
\mbox{}+ \left(1.693 s^2 c^2-0.559 s^4 -1.350 \left(1 - s^4\right)
\right) {1 \over x}
 \right) ~.
\eeqa
The full list of changes to the electroweak observables that we use in
our fits for the heavy case non-commuting ETC model appears in
Appendix A.

Finally, we can combine previously obtained expressions for the masses
of the heavy $W$ boson and the top quark to constrain $x$.  The
analysis of the mass matrix shows that the mass of the heavy $W$
gauge boson is
\beq
M^H_{W} \approx {\sqrt{x}\over sc} M_W~.
\label{thison}
\eeq
If we combine this equation with the estimate of the top mass
given in equation (\ref{mt}) , we find
\beq
M^H_{W} \approx {210\ {\rm GeV}\over sc} \left({175\ {\rm GeV}\over
m_t}\right)^{1\over 2} \left({3 f_Q\over v}\right)^{3\over 2}
\left({u\over f}\right) ~.
\eeq
The last two factors on the right hand side of this equation are written
so that they are of the order of, or less than, unity when there is no
fine-tuning of the ETC interactions. Therefore, we find
\beq
M^H_{W} \le {210\ {\rm GeV}\over sc}
\label{wbound}
\eeq
which, together with (\ref{thison}), implies that
\beq
x \le 6.9
\label{xbnd}
\eeq
assuming the absence of strong-ETC fine-tuning or an alternative
mechanism for generating the top mass.  As we will see in section 5,
these bounds will not be satisfied: some fine-tuning will be required.


\section{Weak Boson Mixing: Light Case}
\label{sec:light}
\setcounter{equation}{0}

The four-fermion operators induced by ETC gauge boson exchange produce
the same corrections in both the heavy and the light case. The
differences between the two cases arises in the low energy
weak interactions and in how the breaking of
$SU(2)_L$ mixes $Z_1$ and $Z_2$, and $W^{\pm}_1$ and
$W^{\pm}_2$.  In the light case, retaining the standard relation
between the $\mu$ decay rate and $G_F$ yields
\beq
\sqrt{2}G_F = {1\over v^2}~.
\eeq
The charged-current four-fermion weak interactions are of the form
\beq
2\sqrt{2} G_F\left( j_l + j_h \right)^2 + {2 \over u^2} j_h^2
\label{wccl}
\eeq
while the low-energy neutral-current interactions can be written
\beq
2\sqrt{2} G_F \, (j_{3l} + j_{3h} -\, j_{em}\sin^2\theta)^2  +
{2\over \,u^2}\, (j_{3h} - s^2 \sin^2\theta j_{em})^2~.
\label{ncplusl}
\eeq
Note that this time the charged-current weak interactions contain no
new $j_l j_h$ term, so that the weak decays of third-generation
fermions will not be altered from their standard rates (see
e.g. Appendix B, equation (B.22)).

An analysis of weak gauge boson mixing shows that the mass-squared
matrix for the $Z_1$ and $Z_2$ is:
\beq
M_Z^2=\left({e v \over {2 \st}} \right)^2\,
\pmatrix{{1\over \cos^2\theta}&
{ c\over s\cos\theta}\cr{ c\over s\cos\theta}&
{x\over s^2c^2}+{c^2\over s^2}\cr}.
\eeq
The mass-squared matrix for $W_1$ and $W_2$ is obtained by
setting $\ct = 1$ in the above matrix.
We again diagonalize the $W$ and $Z$ mass matrices in the limit of
large $x$ and find
\beq
M_W^2 \approx
\left({\pi \alpha_{em} \over{\sqrt{2} G_F \sin^2\theta}}\right)
\left(1 - {c^4\over x}\right),
\eeq
\beq
W^L\approx W_1-{c^3 s \over {x}}\,W_2
\eeq
and
\beq
M_Z^2 \approx
\left({\pi \alpha_{em} \over{\sqrt{2} G_F \sin^2\theta\cos^2\theta}}\right)
\left(1 - {c^4\over x}\right),
\eeq
\beq
Z^L\approx Z_1-{c^3 s \over {x\cos\theta}}\,Z_2 ~.
\label{dmzgfi}
\eeq
Once again, the custodial isospin relation for the $W$ and $Z$ masses
is preserved, to this order in $1/x$.

Thus, in this case the weak gauge boson mixing produces the following
shift from the standard model $Z$ couplings:
\beq
\delta g^f_L =  - {{c^3 s}\over x}
\left({c \over s}\, T_{3l} - {s \over c} \,T_{3h}\right)~.
\eeq
The difference in the relationship between $G_F$, $M_Z$,
and $\sin^2\theta$ in equation (\ref{dmzgfi}) and the corresponding
relationship in the standard model must also be taken into
account\footnote{In this case, using the notation of
Ref. \protect\cite{fit}, we can account for the effect by the
replacement $\Delta_e + \Delta_\mu = -c^4 / x$.}.

As in the heavy case, the predicted values of electroweak observables are
altered from those given by the standard model \cite{unun,fit}.  For
example, we find that the $W$ mass is altered as follows:
\begin{equation}
 M_W = \left( M_W \right)_{SM} \left(1
+ 0.213 \,c^4  {1 \over x} \right) ~.
\end{equation}
Likewise, one finds that the total width of the $Z$ is changed
to:
\beqa
 \Gamma_Z &=& \left( \Gamma_Z \right)_{SM} \left(1 -0.707 \delta g_L^b -0.144
\delta g_L^\tau + 0.268 \delta g_L^{\nu_\tau}\right.\nonumber\\
 & &\left.\,\,\,\,\,\,\,\,\,\,\,\,\,\,\,\,\,\,\,\,\,\,\,\,\,\,
\mbox{}  + \left(-0.343 c^4 + 0.559 s^2 c^2 \right) {1 \over x} \right) ~.
\eeqa
The full list of changes to the electroweak observables used in our
fits for the light case non-commuting ETC model is given in Appendix
B.

\section{Comparison with Data}
\label{sec:data}
\setcounter{equation}{0}

Using the current experimental values of the electroweak observables
and the corresponding best-fit {\it standard model} predictions, we
have used the equations given in Appendices A and B to fit the heavy
case and light case non-commuting ETC model predictions to the data.
Our analysis determines how well each model fits the data, and whether
fine-tuning of the ETC interactions is required.

Even before performing multi-variable fits to the precision
electroweak data, we can place a constraint on this class of models.
We must require that technicolor coupling at the ETC scale,
$f$, not be so strong as to exceed the ``critical" value at which
technifermion chiral symmetry breaking occurs. If we use the results
of the gap-equation analysis of chiral symmetry breaking in the
``rainbow" approximation
\cite{ladder} to estimate this value, we find that \cite{NCETC}
\beq
c^2 > 0.03 \left({N^2-1 \over 2N} \right) ~,
\label{s2bound}
\eeq
where we have assumed that the technifermions form a fundamental
representation, $N$, of an $SU(N)$ technicolor group.

Before describing the details of the fit, we discuss higher-order
corrections. Beyond tree-level, the predictions of the
standard or non-commuting ETC models depend on the values of
$\alpha_s(M_Z)$ and the top-quark mass $m_t$. Given the success of the
standard model, we expect that, for the allowed range of $s^2$, $1/x$,
and the various $\delta g$'s, the changes in the predicted values of
physical observables due to radiative corrections in the standard
model or non-commuting ETC models will be approximately the same {\it
for the same values of $\alpha_s(M_Z)$ and $m_t$}.

The best-fit standard model predictions which we use \cite{langacker}
are based on a top quark mass of 173 GeV (taken from a fit to
precision electroweak data) which is consistent with
the range of masses ($176\pm13$ GeV) preferred by
CDF and consistent with D0 \cite{cdf}.

The treatment of $\alpha_s(M_Z)$ is more problematic: the LEP
determination for $\alpha_s(M_Z)$ comes from a {\it fit} to
electroweak observables {\it assuming} the validity of the standard
model. For this reason it is important \cite{alphaZbb} to understand
how the bounds vary for different values of $\alpha_s(M_Z)$. We
present results for bounds on $s^2$, $1/x$, and the $\delta g$'s, both
for $\alpha_s(M_Z) = 0.124$ (which is the LEP best-fit value assuming
the standard model is correct \cite{langacker}) and for
$\alpha_s(M_Z)=0.115$ as suggested by recent lattice results
\cite{lattice} and deep-inelastic scattering \cite{langacker,deep}.
To the accuracy to which we work, the $\alpha_s$ dependence of the
standard model predictions only appears in the $Z$ partial widths and
we use \cite{langacker}
\beq
\Gamma_q = \Gamma_q|_{\alpha_s=0}\left(1+ {{\alpha_s}\over{\pi}}
+ 1.409 \left({{\alpha_s}\over{\pi}}\right)^2 -
12.77\left({{\alpha_s}\over{\pi}}\right)^3 \right)
\eeq
to obtain the standard model predictions for $\alpha(M_Z)=0.115$.

We have performed a global fit for the parameters of the non-commuting
ETC model ($s^2$, $1/x$, and the $\delta g$'s) to all precision
electroweak data: the $Z$ line shape, forward backward asymmetries,
$\tau$ polarization, and left-right asymmetry measured at LEP and SLC;
the $W$ mass measured at FNAL and UA2; the electron and neutrino
neutral current couplings determined by deep-inelastic scattering; the
degree of atomic parity violation measured in Cesium; and the ratio of
the decay widths of $\tau \to \mu \nu \bar\nu$ and $\mu\to e \nu \bar
\nu$.  Essentially we find that, while both the heavy and light cases
provide a better fit to the data than the standard model, the light
case can reproduce the data with much smaller gauge boson masses, and
hence is of more phenomenological interest.

In Table \ref{Pred} we compare the predictions of the standard model
and the non-commuting ETC model (with particular values of $1/x$ and
$s^2$) with the experimental values.  For $s^2$, we have chosen a
value of 0.97 which saturates our bound (\ref{s2bound}),
since this conforms to our expectation that the ETC gauge coupling is
quite strong.  For $1/x$, in the heavy case we show the best fit
value of $1/x =0.0027$ or equivalently $M_W^H= 9$ TeV.
The choice of a particular value of $1/x$ for the light case is fairly
arbitrary.  We do not show the best fit case, since it lies the
unphysical region of negative $x$ (the fit gives $1/x = -0.17 \pm
0.75$).  However, since the fit is fairly insensitive to the value of
$1/x$ (i.e. the uncertainty in $1/x$ is large) there is a
substantial range of values for $1/x$ which provide a good fit to the
data. For illustration we have chosen the value $1/x = 0.055$ which
corresponds to $M_W^H= 2$ TeV.

\begin{table}[htbp]
\begin{center}
\begin{tabular}{|c|l|l|l|l|}\hline\hline
Quantity & Experiment & SM & ${\rm ETC}_{\rm heavy}$ & ${\rm ETC}_{\rm light}$
\\\hline \hline
$\Gamma_Z$ & 2.4976 $\pm$ 0.0038 & 2.4923 & 2.4991 & 2.5006 \\
$R_e$ & 20.86 $\pm$ 0.07 & 20.73 & 20.84 & 20.82 \\
$R_\mu$ & 20.82 $\pm$ 0.06 & 20.73 & 20.84 & 20.82 \\
$R_\tau$ & 20.75 $\pm$ 0.07 & 20.73 & 20.74 & 20.73 \\
$\sigma_h$ & 41.49 $\pm$ 0.11 & 41.50 & 41.48 & 41.40 \\
$R_b$ & 0.2202 $\pm$ 0.0020 & 0.2155 & 0.2194 & 0.2188 \\
$A_{FB}^e$ & 0.0156 $\pm$ 0.0034 & 0.0160 & 0.0159 & 0.0160 \\
$A_{FB}^\mu$ & 0.0143 $\pm$ 0.0021 & 0.0160 & 0.0159 & 0.0160 \\
$A_{FB}^\tau$ & 0.0230 $\pm$ 0.0026 & 0.0160 & 0.0164 & 0.0164 \\
$A_{\tau}(P_\tau)$ & 0.143 $\pm$ 0.010 & 0.146 & 0.150 & 0.150 \\
$A_{e}(P_\tau)$ & 0.135 $\pm$ 0.011 & 0.146 & 0.146 & 0.146 \\
$A_{FB}^b$ & 0.0967 $\pm$ 0.0038 & 0.1026 & 0.1026 & 0.1030 \\
$A_{FB}^c$ & 0.0760 $\pm$ 0.0091 & 0.0730 & 0.0728 & 0.0730 \\
$A_{LR}$ & 0.1637 $\pm$ 0.0075 & 0.1460 & 0.1457 & 0.1460 \\
$M_W$ & 80.17 $\pm$ 0.18 & 80.34 & 80.34 & 80.34 \\
$M_W/M_Z$ & 0.8813 $\pm$ 0.0041 & 0.8810 & 0.8810 & 0.8810 \\
$g_L^2(\nu N \rightarrow \nu X)$ & 0.3003 $\pm$ 0.0039 & 0.3030 & 0.3026 &
0.3030  \\
$g_R^2(\nu N \rightarrow \nu X)$ & 0.0323 $\pm$ 0.0033 & 0.0300 & 0.0301 &
0.0300  \\
$g_{eA}(\nu e \rightarrow \nu e)$ & -0.503 $\pm$ 0.018 & -0.506 & -0.506 &
-0.506 \\
$g_{eV}(\nu e \rightarrow \nu e)$ & -0.025 $\pm$ 0.019 & -0.039 & -0.038 &
-0.039 \\
$Q_W(Cs)$ & -71.04 $\pm$ 1.81 & -72.78 &  -72.78 & -72.78 \\
$R_{\mu \tau}$ & 0.9970 $\pm$ 0.0073 & 1.0 & 0.9946 & 1.0  \\
\hline\hline
\end{tabular}
\end{center}
\caption{Experimental \protect\cite{langacker,blondel,taudec} and
predicted values of
electroweak observables for the standard model and non-commuting ETC
model (heavy and light cases) for $\alpha_s(M_Z)=0.115$, and $s^2=0.97$.
For the heavy case $1/x$ is allowed to assume the best-fit value of
0.0027; for the light case, $1/x$ is set to  $0.055$. The standard model values
correspond to the best-fit values (with $m_t=173$ GeV, $m_{\rm Higgs}
= 300$ GeV) in \protect\cite{langacker}, corrected for the change in
$\alpha_s(M_Z)$, and the revised extraction \protect\cite{swartz} of
$\alpha_{em}(M_Z)$.}
\label{Pred}
\end{table}

\subsection{The Light Case}
\label{subsec:lightt}

We will first discuss the light case in some detail.  We have fit the
precision electroweak data to the expressions in Appendix B, allowing
$s^2, 1/x,$ and the $\delta g$'s to vary.  Figure 1 summarizes the
fits by displaying the 95\% and 68\% confidence level lower bounds
(solid and dotted lines) on the heavy $W$ mass ($M^H_{W}$) for different
values of $s^2$ (using $\alpha_s(M_Z)=0.115$ as before).  The plot was
created as follows: for each value of $s^2$ we fit to the three
independent parameters ($\delta g_L^b$, $\delta g_L^\tau = \delta
g_L^{\nu_\tau}$, and $1/x$); we then found the lower bound on $x$ and
translated it into a lower bound on the heavy $W$ mass.
Note that for $s^2 > 0.85$, the 95\% confidence level
allows the heavy $W$ gauge boson to be as light as $400$ GeV.

As mentioned previously, the predictions of the non-commuting ETC
model in the light case are given in Table \ref{Pred}.  The question
remains as to how well this model fits the precision data.  The
summary of the quality of the fit is given in Table
\ref{lightFit115}. The table shows the fit to the standard model for
comparison; as a further benchmark we have included a fit to purely
oblique corrections (the $S$ and $T$ parameters) \cite{ST}.  The
percentage quoted in the Table is the probability of obtaining a
$\chi^2$ as large or larger than that obtained in the fit, for the
given number of degrees of freedom (df), assuming that the model is
correct.  Thus a small probability corresponds to a poor fit, and a
bad model.  The $SM+S,T$ fit demonstrates that merely having more
parameters is not sufficient to ensure a better fit.

\begin{table}[htbp]
\begin{center}
\begin{tabular}{|l||c||c||c||c|}\hline\hline
Model &  $\chi^2$&df&$\chi^2/{\rm df}$  & probability \\ \hline\hline
SM & $33.8$ & $22$& $1.53$ & $5\%$ \\ \hline
SM+$S$,$T$  &$32.8$ & $20$ & $1.64$ & $4\%$ \\ \hline
${\rm ETC}_{\rm light}$  &$22.6$ & $20$ & $1.13$ & $31\%$ \\ \hline
\hline
\end{tabular}
\end{center}
\caption{The best fits for the standard model, beyond the standard model
allowing $S$  and $T$
to vary, and the non-commuting ETC model (light case).  The inputs are:
$\alpha_s(M_Z)=0.115$,
 $1/x=0.055$, and $s^2=0.97$. $\chi^2$ is the sum
of the squares of the difference between prediction and experiment,
divided by the error.}
\label{lightFit115}
\end{table}

The best fit values for the shifts in
the $Z$ couplings (corresponding to
Tables \ref{Pred} and \ref{lightFit115}, i.e. the light case with
$\alpha_s(M_Z)= 0.115$) are:
\begin{eqnarray}
\delta g_L^b &=& -0.0035 \pm 0.0015 \nonumber\\
\delta g_L^\tau &=& \delta g_L^{\nu_\tau} = -0.0002\pm 0.0009 ~.
\label{ZShiftLight115}
\end{eqnarray}
The 90\% confidence region for this fit is shown as the solid line in Figure 2.
{}From equation (\ref{ZShiftLight115}) and Figure 2 we see that the
data prefers a significant shift in the
$Z b \overline{b}$ coupling.

Using the values above and equation (\ref{dgLb}) we can obtain
a 95\% limit on the scale $f$:
\beq
f > \xi \cdot 123 \,{\rm GeV}.
\eeq
Using equation (\ref{fine}) we see that we require a fine-tuning of
the ETC coupling of order:
\beq
{{4 M^2}\over{g^2f^2}} \approx 52{\rm \%}\,\,
{\left(f_Q \over {30 \,{\rm GeV}}\right)^3}
{\left({123 \,{\rm GeV} \over f }\right)^2}
{\left({175 \,{\rm GeV} \over m_t }\right)}
\, .
\eeq
{}From Tables \ref{Pred} and \ref{lightFit115}
we see that because the theory accommodates
changes in the $Z$ partial widths, the non-commuting ETC model gives a
significantly better fit to the experimental data than the standard
model does, even after taking into account that in the fitting
procedure the non-commuting ETC model has two extra parameters.
In particular the non-commuting ETC model predicts values for
$\Gamma_Z$, $R_e$, $R_\mu$, $R_\tau$,  and $R_b$ that are closer to
experiment than those predicted by the standard model.

For comparison we have also performed the fits using $\alpha_s(M_Z)=0.124$;
the quality of the fit is summarized in Table \ref{lightFit124}.
The best fit values for the shifts in the $Z$ couplings (corresponding to
Table \ref{lightFit124}, i.e. the light case with $\alpha_s(M_Z)= 0.124$) are:
\begin{eqnarray}
\delta g_L^b &=& -0.0013 \pm 0.0015 \nonumber\\
\delta g_L^\tau &=& \delta g_L^{\nu_\tau} = -0.0003\pm 0.0009  ~.
\end{eqnarray}
The 90\% confidence region for this fit is shown as the dashed line in
Figure 2.  We find that, while the standard model fit improves for a
larger value of $\alpha_s(M_Z)$, the light case of the non-commuting
ETC model remains a somewhat better fit.

\begin{table}
\begin{center}
\begin{tabular}{|l||c||c||c||c|}\hline\hline
Model &  $\chi^2$&df&$\chi^2/{\rm df}$  & probability \\ \hline\hline
SM & $27.8$ & $21$& $1.33$ & $15\%$ \\ \hline
SM+$S$,$T$  &$27.7$ & $20$ & $1.38$ & $12\%$ \\ \hline
${\rm ETC}_{\rm light}$  &$25.0$ & $20$ & $1.25$ & $20\%$ \\ \hline
\hline
\end{tabular}
\end{center}
\caption{The best fits for the standard model, beyond the standard model
allowing $S$ and $T$
to vary, and the non-commuting ETC model (light case).  The inputs are:
$\alpha_s(M_Z)=0.124$,
 $1/x=0.055$, and $s^2=0.97$.}
\label{lightFit124}
\end{table}

\subsection{The Heavy Case}
\label{subsec:heavyy}

For the heavy case, we have
fit the precision electroweak data to the expressions in Appendix
A, allowing $s^2$, $1/x$ and the $\delta g$'s to vary.
Our results are summarized in Figure 3, which displays the 95\%
and 68\% confidence lower bounds on the heavy $W$ boson
(solid and dotted lines) as a
function of $s^2$.   For comparison, we show
the upper bound (dashed line) on the heavy $W$ mass that is supplied
by equation (\ref{wbound}) in the absence of fine-tuned ETC
interactions.  No region of the plot satisfies both bounds,
i.e. the model must be fine tuned in order to produce the top
mass and agree with precision measurements.
Allowing for the possibility of some fine tuning,
the lowest possible heavy $W$ mass at the
95\% confidence level  is roughly 1.6 TeV, for $0.7< s^2 <0.8$.
This corresponds to $f > 2$ TeV, and hence a tuning of order
\beq
{{4 M^2}\over{g^2f^2}} \approx 14{\rm \%}\,\,
{\left(f_Q \over {125 \, {\rm GeV}}\right)^3}
{\left({2 \,{\rm TeV} \over f }\right)^2}
{\left({175 \,{\rm GeV} \over m_t }\right)}
\, .
\eeq

A summary of the quality of the global fit to the precision
electroweak data is given in Table \ref{heavyFit115}.
We conclude that the heavy case also gives a good fit to
the data, but the masses of the new gauge boson are
substantially heavier than in the light case.

\begin{table}[htb]
\begin{center}
\begin{tabular}{|l||c||c||c||c|}\hline\hline
Model &  $\chi^2$&df&$\chi^2/{\rm df}$  & probability \\ \hline\hline
SM & $33.8$ & $22$& $1.53$ & $5\%$ \\ \hline
SM+$S$,$T$  &$32.8$ & $20$ & $1.64$ & $4\%$ \\ \hline
${\rm ETC}_{\rm heavy}$  &$20.7$ & $19$ & $1.09$ & $36\%$ \\ \hline
\hline
\end{tabular}
\end{center}
\caption[best fits]{The best fits for the standard model, the
standard model plus extra oblique corrections, and the non-commuting
ETC model (heavy case).  The inputs are $\alpha_s(M_Z)=0.115$, and
(for the ETC model) $s^2=0.97$.  }
\label{heavyFit115}
\end{table}

The best fit values for the parameters (corresponding to
Tables \ref{Pred} and \ref{heavyFit115}, i.e. the heavy case with
$\alpha_s(M_Z)= 0.115$) are:
\begin{eqnarray}
1/x &=& 0.0027  \pm 0.0093 \nonumber \\
\delta g_L^b &=& -0.0064 \pm 0.0074 \\
\delta g_L^\tau &=& \delta g_L^{\nu_\tau} = -0.0024\pm 0.0055 ~.\nonumber
\label{ZShiftHeavy115}
\end{eqnarray}



\section{Conclusions}
\label{sec:concl}
Surprisingly we have found that both cases of the non-commuting ETC
model can give a significantly better fit to experimental data than
the standard model, even when one takes into account the fact that the
ETC model has extra parameters that are fit to data.  Part of this
relative success of course is due to the fact that the standard model
does not fit the data very well (contrary to current folklore),
especially if $\alpha_s(M_Z)=0.115$.  One might object that the
non-commuting ETC model is not an aesthetically pleasing model.
Furthermore both the heavy and light cases required some fine-tuning.
Nonetheless, our results provide a significant existence proof: there
is at least one ETC model that can fit the precision electroweak data
{\it better} than the standard model.  The light case of this
non-commuting ETC model is especially interesting since the heavy
gauge bosons can be lighter than 1 TeV, and hence could potentially be
directly produced at foreseeable accelerators.


\newpage
\centerline{\bf Acknowledgments}
\vspace{12pt}

J.T. thanks W. Marciano and W. Williams  for helpful conversations
regarding tau decays and statistics respectively.
We thank K. Lane for comments on the manuscript.

R.S.C. acknowledges the support of an NSF Presidential Young
Investigator Award, and a DOE Outstanding Junior Investigator Award.
E.H.S.  acknowledges the support of an NSF Faculty Early Career Development
(CAREER) award. {\em This work was supported in part by the National
Science Foundation under grants PHY-9057173 and PHY-9501249, and by
the Department of Energy under grant DE-FG02-91ER40676.}

\appendix
\section{Appendix: Equations for heavy case}
\label{sec:heavyappx}
\setcounter{equation}{0}

The full list of corrections \cite{NCETC,fit} to standard
model  predictions in the `heavy case' of non-commuting ETC is:
\beqa
 \Gamma_Z &=& \left( \Gamma_Z \right)_{SM} \left(1 -0.707 \delta g_L^b
-0.144  \delta g_L^\tau + 0.268 \delta g_L^{\nu_\tau}\right.\nonumber\\
 & &\left.\,\,\,\,\,\,\,\,\,\,\,\,\,\,\,\,\,\,\,\,\,\,\,\,\,\,
\mbox{}+ \left(1.693 s^2 c^2-0.559 s^4 -1.350 \left(1 - s^4\right)
\right)  {1 \over x}
 \right)
\eeqa
\begin{equation}
 R_e = \left( R_e \right)_{SM}
\left(1 -1.01 \delta g_L^b  + \left(-0.313 s^2 c^2-0.505 s^4 -0.260
\left(1 - s^4\right) \right) {1 \over x} \right)
\end{equation}
\begin{equation}
 R_\mu = \left( R_\mu \right)_{SM} \left(1 -1.01 \delta g_L^b  +
\left(-0.313 s^2 c^2-0.505 s^4 -0.260 \left(1 - s^4\right) \right)
{1 \over x} \right)
\end{equation}
\beqa
 R_\tau &=&
 \left( R_\tau \right)_{SM} \left(1 -1.01 \delta g_L^b  +
4.290 \delta g_L^\tau \right.\nonumber\\
 & &\left.\,\,\,\,\,\,\,\,\,\,\,\,\,\,\,\,\,\,\,\,\,\,\,\,\,
\mbox{}+ \left(1.832 s^2 c^2 + 1.640 s^4 -0.260 \left(1 - s^4\right)
\right)  {1 \over x} \right)
\eeqa
\beqa
 \sigma_h &=& \left( \sigma_h \right)_{SM} \left(1 + 0.404 \delta g_L^b   +
0.288 \delta g_L^\tau -0.536 \delta g_L^{\nu_\tau} \right.\nonumber\\
 & &\left.\,\,\,\,\,\,\,\,\,\,\,\,\,\,\,\,\,\,\,\,\,\,\,\,\,
\mbox{}+ \left(0.591 s^2 c^2 + 0.614 s^4  + 0.022
\left(1 - s^4\right) \right) {1 \over x}\right)
\eeqa
\begin{equation}
 R_b = \left( R_b \right)_{SM} \left(1 -3.56 \delta g_L^b  +
\left(-1.832 s^2 c^2-1.780 s^4  + 0.059 \left(1 - s^4\right) \right)
{1 \over x} \right)
\end{equation}
\begin{equation}
 A_{FB}^e = \left( A_{FB}^e \right)_{SM}  + \left(0.430 s^2 c^2-0.614
\left(1 - s^4\right) \right) {1 \over x}
\end{equation}
\begin{equation}
 A_{FB}^\mu = \left( A_{FB}^\mu \right)_{SM}  + \left(0.430 s^2 c^2-0.614
\left(1 - s^4\right) \right) {1 \over x}
\end{equation}
\begin{equation}
 A_{FB}^\tau = \left( A_{FB}^\tau \right)_{SM} -0.430 \delta g_L^\tau  +
\left(0.215 s^2 c^2-0.215 s^4 -0.614 \left(1 - s^4\right) \right)
{1 \over x}
\end{equation}
\begin{equation}
 A_{\tau}(P_\tau) = \left( A_{\tau}(P_\tau) \right)_{SM}
-3.610 \delta g_L^\tau+
\left(-1.805 s^4 -2.574 \left(1 - s^4\right) \right) {1 \over x}
\end{equation}
\begin{equation}
 A_{e}(P_\tau) = \left( A_{e}(P_\tau) \right)_{SM}  + \left(1.805 s^2 c^2-2.574
\left(1 - s^4\right) \right) {1 \over x}
\end{equation}
\begin{equation}
 A_{FB}^b = \left( A_{FB}^b \right)_{SM} -0.035 \delta g_L^b  + \left(1.269 s^2
c^2-0.017 s^4 -1.828 \left(1 - s^4\right) \right) {1 \over x}
\end{equation}
\begin{equation}
 A_{FB}^c = \left( A_{FB}^c \right)_{SM}  + \left(1.003 s^2 c^2-1.433 \left(1 -
s^4\right) \right) {1 \over x}
\end{equation}
\begin{equation}
 A_{LR} = \left( A_{LR} \right)_{SM}  + \left(1.805 s^2 c^2-2.574 \left(1 -
s^4\right) \right) {1 \over x}
\end{equation}
\begin{equation}
 M_W = \left( M_W \right)_{SM} \left(1
-0.213 \left(1 - s^4\right)  {1 \over x} \right)
\end{equation}
\begin{equation}
 M_W/M_Z = \left( M_W/M_Z \right)_{SM} \left(1
-0.213 \left(1 - s^4\right)  {1 \over x} \right)
\end{equation}
\begin{equation}
 g_L^2(\nu N \rightarrow \nu X) =
\left( g_L^2(\nu N \rightarrow \nu X) \right)_{SM}
 -0.244 \left(1 - s^4\right)  {1 \over x}
\end{equation}
\begin{equation}
 g_R^2(\nu N \rightarrow \nu X) =
\left( g_R^2(\nu N \rightarrow \nu X) \right)_{SM}
+  0.085 \left(1 - s^4\right)  {1 \over x}
\end{equation}
\begin{equation}
 g_{eA}(\nu e \rightarrow \nu e) =
\left( g_{eA}(\nu e \rightarrow \nu e) \right)_{SM}
\end{equation}
\begin{equation}
 g_{eV}(\nu e \rightarrow \nu e) =
\left( g_{eV}(\nu e \rightarrow \nu e) \right)_{SM}
 + 0.656 \left(1 - s^4\right)  {1 \over x}
\end{equation}
\begin{equation}
 Q_W(Cs) = \left( Q_W(Cs) \right)_{SM}  + \left(-21.16\,  c^2 + 1.450
\left(1 - s^4\right) \right) {1 \over x}
\end{equation}
\beq
R_{\mu \tau} \equiv {\Gamma(\tau \to \mu \nu \bar\nu)\over
\Gamma(\mu\to e \nu \bar \nu)} = R_{\mu \tau}^{SM} ( 1 - {2\over{x}})
\eeq

\section{Appendix: Equations for light case}
\label{sec:lightappx}
\setcounter{equation}{0}

We obtain the following corrections \cite{fit,NCETC} to standard model
predictions in the light case of the non-commuting ETC models:
\beqa
 \Gamma_Z &=& \left( \Gamma_Z \right)_{SM} \left(1 -0.707 \delta g_L^b -0.144
\delta g_L^\tau + 0.268 \delta g_L^{\nu_\tau}\right.\nonumber\\
 & &\left.\,\,\,\,\,\,\,\,\,\,\,\,\,\,\,\,\,\,\,\,\,\,\,\,\,\,
\mbox{}  + \left(-0.343 c^4 + 0.559 s^2 c^2 \right) {1 \over x} \right)
\eeqa
\begin{equation}
 R_e = \left( R_e \right)_{SM} \left(1 -1.01 \delta g_L^b  + \left(0.573 c^4 +
0.505 s^2 c^2 \right) {1 \over x} \right)
\end{equation}
\begin{equation}
 R_\mu = \left( R_\mu \right)_{SM} \left(1 -1.01 \delta g_L^b  + \left(0.573
c^4
+ 0.505 s^2 c^2 \right) {1 \over x} \right)
\end{equation}
\begin{equation}
 R_\tau = \left( R_\tau \right)_{SM} \left(1 -1.01 \delta g_L^b + 4.29 \delta
g_L^\tau  + \left(-1.572 c^4-1.640 s^2 c^2 \right) {1 \over x} \right)
\end{equation}
\beqa
 \sigma_h &=& \left( \sigma_h \right)_{SM} \left(1 + 0.404 \delta g_L^b + 0.288
\delta g_L^\tau -0.536 \delta g_L^{\nu_\tau} \right.\nonumber\\
 & &\left.\,\,\,\,\,\,\,\,\,\,\,\,\,\,\,\,\,\,\,\,\,\,\,\,\,
\mbox{}  + \left(-0.613 c^4-0.614 s^2 c^2 \right) {1 \over x} \right)
\eeqa
\begin{equation}
 R_b = \left( R_b \right)_{SM} \left(1 -3.56 \delta g_L^b  + \left(0.129 c^4 +
1.780 s^2 c^2 \right) {1 \over x} \right)
\end{equation}
\begin{equation}
 A_{FB}^e = \left( A_{FB}^e \right)_{SM}
+ 0.184 c^4  {1 \over x}
\end{equation}
\begin{equation}
 A_{FB}^\mu = \left( A_{FB}^\mu \right)_{SM}
+ 0.184 c^4  {1 \over x}
\end{equation}
\begin{equation}
 A_{FB}^\tau = \left( A_{FB}^\tau \right)_{SM} -0.430 \delta g_L^\tau  +
\left(0.399 c^4 + 0.215 s^2 c^2 \right) {1 \over x}
\end{equation}
\begin{equation}
 A_{\tau}(P_\tau) = \left( A_{\tau}(P_\tau) \right)_{SM}
-3.610 \delta g_L^\tau  + \left(2.574 c^4 + 1.805 s^2 c^2 \right) {1 \over x}
\end{equation}
\begin{equation}
 A_{e}(P_\tau) = \left( A_{e}(P_\tau) \right)_{SM}
+ 0.769 c^4  {1 \over x}
\end{equation}
\begin{equation}
 A_{FB}^b = \left( A_{FB}^b \right)_{SM} -0.035 \delta g_L^b
+ \left(0.520 c^4 + 0.161 s^2 c^2 \right) {1 \over x}
\end{equation}
\begin{equation}
 A_{FB}^c = \left( A_{FB}^c \right)_{SM}
+ 0.400 c^4  {1 \over x}
\end{equation}
\begin{equation}
 A_{LR} = \left( A_{LR} \right)_{SM}
+ 0.769 c^4  {1 \over x}
\end{equation}
\begin{equation}
 M_W = \left( M_W \right)_{SM} \left(1
+ 0.213 c^4  {1 \over x} \right)
\end{equation}
\begin{equation}
 M_W/M_Z = \left( M_W/M_Z \right)_{SM} \left(1  +
0.213 c^4  {1 \over x} \right)
\end{equation}
\begin{equation}
 g_L^2(\nu N \rightarrow \nu X) =
\left( g_L^2(\nu N \rightarrow \nu X) \right)_{SM}
-0.529 c^4  {1 \over x}
\end{equation}
\begin{equation}
 g_R^2(\nu N \rightarrow \nu X) =
\left( g_R^2(\nu N \rightarrow \nu X) \right)_{SM}  +
0.850 c^4  {1 \over x}
\end{equation}
\begin{equation}
 g_{eA}(\nu e \rightarrow \nu e) =
\left( g_{eA}(\nu e \rightarrow \nu e) \right)_{SM}  +
0.500 c^4  {1 \over x}
\end{equation}
\begin{equation}
 g_{eV}(\nu e \rightarrow \nu e) =
\left( g_{eV}(\nu e \rightarrow \nu e) \right)_{SM}
-0.156 c^4  {1 \over x}
\end{equation}
\begin{equation}
 Q_W(Cs) = \left( Q_W(Cs) \right)_{SM}  +
95.05 c^4  {1 \over x}
\end{equation}
\beq
R_{\mu \tau} \equiv {\Gamma(\tau \to \mu \nu \bar\nu)\over
\Gamma(\mu\to e \nu \bar \nu)} = R_{\mu \tau}^{SM}
\eeq


\newpage
\centerline{\bf Figure Captions}
\vspace{24pt}
\vspace{24pt}
Figure 1. The solid line is the 95\%
confidence lower bound for $M^H_W$
as a function of $s^2$  for the light case
(using $\alpha_s(M_Z)=0.115$). The dotted line
is the 68\% confidence lower bound.

\vspace{24pt}
Figure 2.
The 90\% confidence region for ETC induced shifts in $Z$ couplings for
the light case (using $1/x = 0.028$ and $s^2 = 0.97$)
for $\alpha_s(M_Z)=0.115$ (solid line), and
$\alpha_s(M_Z)=0.124$ (dashed line).  Note that the standard model
prediction (the origin) is excluded for $\alpha_s(M_Z)=0.115$.
\vspace{24pt}

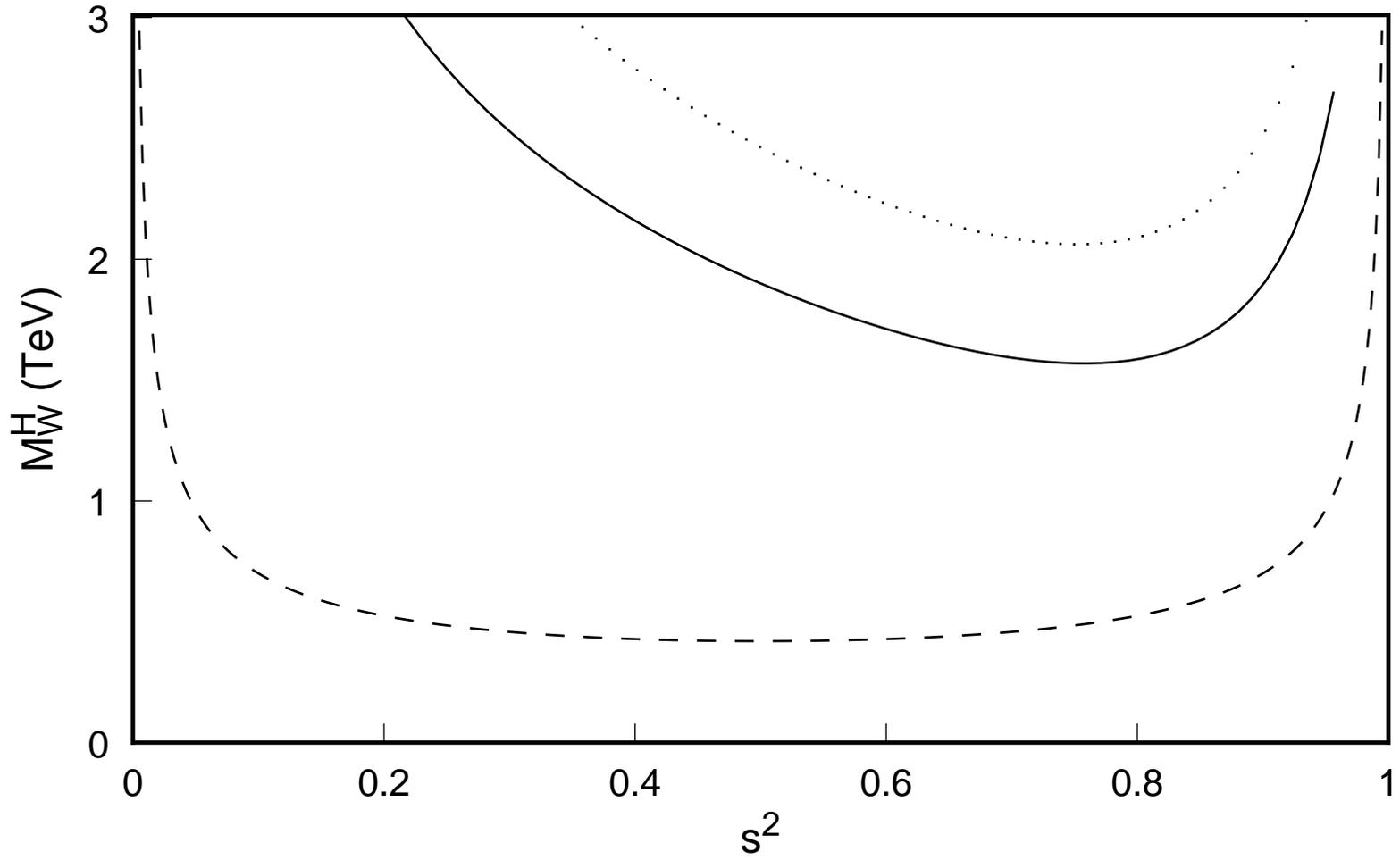
Figure 3.  The solid line is the 95\%
confidence lower bound for $M^H_W$
as a function of $s^2$  for the heavy case
(using $\alpha_s(M_Z)=0.115$).   The dotted line
is the 68\% confidence lower bound.
The dashed line is the upper
bound  on  $M^H_W$ (in the absence of ETC fine-tuning).
Note that there is no overlap region.

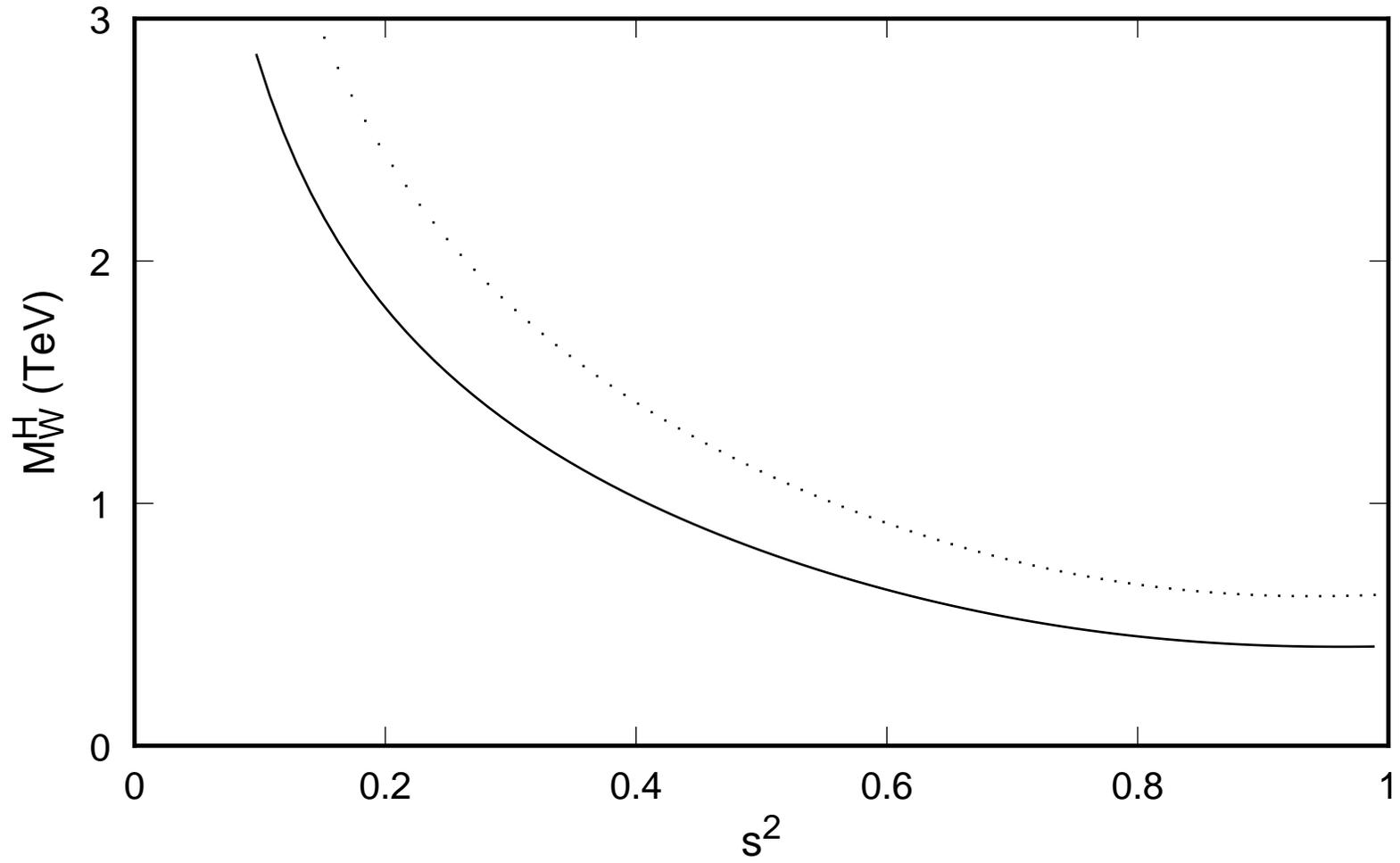
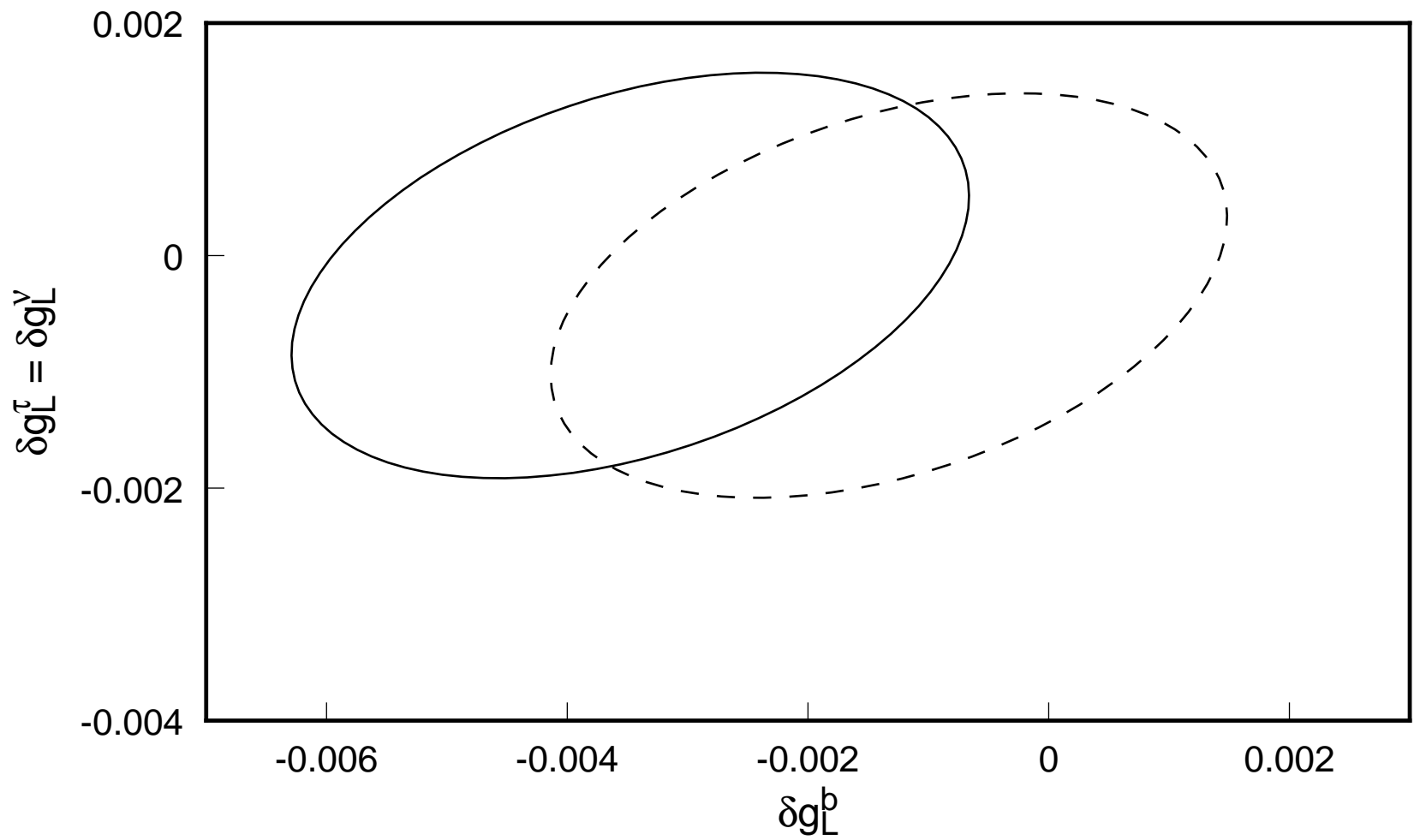
\end{document}